\documentclass{aip-cp}

\usepackage[numbers]{natbib}
\usepackage{rotating}
\usepackage{graphicx}
\usepackage{mathrsfs}
\usepackage{amssymb}
\usepackage{graphicx}
\usepackage[normalem]{ulem}
\usepackage{color}
\usepackage{bm}
\usepackage{longtable}
\usepackage{times}

\usepackage{natbib}
\usepackage{isotope}

\newcommand{\fmiq}{\,\mathrm{fm}^{-3}}

\newcommand{\km}{\,\mathrm{km}}

\newcommand{\beq}{\begin{equation}}
\newcommand{\eeq}{\end{equation}}
\newcommand{\beqa}{\begin{eqnarray}}
\newcommand{\eeqa}{\end{eqnarray}}

\renewcommand\sout{\bgroup \color{red} \ULdepth=-.5ex \ULset}

\renewcommand{\rm}[1]{\textrm{#1}}

\begin{document}


\title{Constraining the properties of dense matter and neutron stars by combining nuclear physics and gravitational waves from GW170817}

\author[aff1]{I. Tews}
\author[aff2]{J. Margueron\corref{cor1}}
\corresp[cor1]{Corresponding author and speaker: j.margueron@ipnl.in2p3.fr}
\author[aff3,aff4]{S. Reddy}
\affil[aff1]{Theoretical Division, Los Alamos National Laboratory, Los Alamos, NM 87545, USA}
\affil[aff2]{Institut de Physique Nucl\'eaire de Lyon, CNRS/IN2P3, Universit\'e de Lyon, Universit\'e Claude Bernard Lyon 1, F-69622 Villeurbanne Cedex, France}
\affil[aff3]{Institute for Nuclear Theory, University of Washington, Seattle, WA 98195-1550, USA}
\affil[aff4]{JINA-CEE, Michigan State University, East Lansing, MI, 48823, USA}
\maketitle

\begin{abstract}
Gravitational waves from neutron-star mergers are expected to provide stringent constraints on the structure of neutron stars. 
At the same time, recent advances in nuclear theory have enabled reliable calculations of the low density equation of state using effective field theory based Hamiltonians and advanced techniques to solve the quantum many-body problem. In this paper, we address how the first observation of gravitational waves from GW170817 can be combined with modern calculations of the equation of state to extract useful insights about the equation of state of matter encountered inside neutron stars. We analyze the impact of various uncertainties and we show that the tidal deformability extracted from GW170817 is compatible, while less constraining, than modern nuclear physics knowledge.
 \end{abstract}


\section{GW170817: the first observation of gravitational waves from binary neutron star merger}

We are living an exciting time for the understanding of dense matter properties in compact stars. Both gravitational wave detectors and accurate X-ray observations are expected to bring decisive constrains that will hopefully help to answering many of the present questions, such as the equation of state (EoS) of dense matter, the onset of phase transitions, and the composition of matter at very high density.
Neutron-star merger events, for instance, simultaneously emit gravitational waves (GWs) and electromagnetic (EM) signals, from gamma-rays, X-rays, optical, infrared, to radio waves, and neutrinos. The first observation of a NS merger by the LIGO and Virgo (LV) interferometers, GW170817 in the GW spectrum, GRB 170817A in the gamma-ray spectrum, and AT~2017gfo in the electromagnetic (EM) spectrum, was made on August 17, 2017, and in the weeks thereafter~\cite{TheLIGOScientific:2017qsa,GBM:2017lvd,Monitor:2017mdv,Abbott:2018wiz}. Triggered by the Fermi and Integral telescopes~\cite{Monitor:2017mdv,Savchenko:2017ffs}, this observation provided detailed spectral and temporal features both in GWs and EM radiation. Theoretical efforts to interpret this data has provided insights into the production of heavy r-process elements in NS mergers~\cite{Drout:2017ijr}, and constraints on the EOS of dense matter~\cite{Annala:2017llu,Fattoyev:2017jql,Most:2018hfd,Lim:2018bkq,Tews:2018iwm}. NS mergers have the potential to provide detailed information on the properties of the merging compact stars, such as their masses and radii~\cite{Bauswein:2017vtn}, as well as on the properties of the densest baryonic matter to be observed in the universe. Since the O3 run of the Advanced LV interferometers have started on April 1st 2019, for a full year, a large number of new detections of NS mergers will provide even stronger constraints on the EoS of strongly-interacting matter and the r-process. 

The LV collaboration observed the GW signal of GW170817 for about $100 s$ (several 1000 revolutions, starting from 25 Hz) and performed detailed analyses of the wave front~\cite{Abbott:2018wiz}. Because the chirp mass $M_\mathrm{chirp}$, defined as 
\begin{equation}
M_\mathrm{chirp}=\frac{(m_1 m_2)^{3/5}}{(m_1+m_2)^{1/5}}\,,
\end{equation} 
can be extracted from the entire signal, this observation allowed to put tight constraints on it. For GW170817, the LV collaboration precisely determined $M_\mathrm{chirp}= 1.186\pm 0.001 M_{\odot}$.

The extraction of higher-order GW parameters from the wavefront is complicated for several reasons, and one of them is the spin of the pulsars.
In this work, we only investigate the low-spin scenario for two reasons. First, large spins are not expected from the observed galactic binary NS population. Second, because neutron stars spin down over time, low spins are also expected from the extremely long merger time of GW170817 of the order of gigayears. Therefore, the low spin scenario is expected to be the more realistic scenario for binary neutron-star mergers such as GW170817.
The above mentioned problems in the extraction of higher-order parameters lead to weaker constraints on the individual masses of the two component neutron stars in GW170817. With $m_1$ being the mass of the heavier and $m_2$ being the mass of the lighter neutron star in the binary, the mass distribution of the individual stars is typically described in terms of the parameter $q=m_2/m_1$. 
The posterior of the LV collaboration for $q$ by the analytical probability distribution~\cite{Tews:2018kmu,Tews2019}
\begin{equation}
p(q)=\exp \left(-\frac12 v(q)^2 -\frac{c}{2} v(q)^4 \right)\,,\label{eq:massdist}
\end{equation}
where $c=1.83$ and $v(q)=(q-0.89)/0.20$.

The tidal polarizability describes how a neutron star deforms under an external gravitational field, and depends on neutron-star properties as
\begin{eqnarray}
\Lambda &=\frac23 k_2 \left(\frac{c^2}{G} \frac{R}{M}\right)^5\,.
\end{eqnarray}
Here, $k_2$ is the tidal love number, that is computed together with the Tolman-Oppenheimer-Volkoff equations; see, for example, Refs.~\cite{Flanagan2008,Damour2009,Moustakidis:2016sab} for more details.

For neutron-star mergers, the GW signal allows the extraction of the binary tidal polarizability parameter $\tilde{\Lambda}$. This parameter is defined as a mass-weighted average of the individual tidal polarizabilities, 
\begin{equation}
\tilde{\Lambda}~=~\frac{16}{13} \left[\frac{(m_1+12m_2)m_1^4\Lambda_1 }{m_{\mathrm{tot}}^5}+ \frac{(m_2+12m_1)m_2^4\Lambda_2 }{m_{\mathrm{tot}}^5}\right]\,.
\end{equation}
As already discussed, the extraction of the binary tidal polarizability suffers from increased uncertainties, due to its importance only during the last few orbits~\cite{Flanagan2008,Damour2009} and correlations among the parameters. In the initial publication of the LV collaboration~\cite{TheLIGOScientific:2017qsa}, the constraint on $\tilde{\Lambda}\leq 800$ was reported with 90\% confidence (corrected to be $\tilde{\Lambda}\leq 900$ in Ref.~\cite{Abbott:2018wiz}). This analysis, however, was very general and did not assume both objects in the binary system to have the same EoS. Several reanalyses have since improved this constraint. Assuming that both compact objects were neutron stars governed by the same EoS, Ref.~\cite{De:2018uhw} used polytropic EoS models and a Bayesian parameter estimation with additional information on the source location from EM observations to derive limits on $\tilde{\Lambda}$ for different prior choices for the component masses: for uniform priors the reported 90\% confidence interval was $\tilde{\Lambda}=84-642$, for a component mass prior informed by radio observations of Galactic double neutron stars the result was $\tilde{\Lambda}=94-698$, and for a component mass prior informed by radio pulsars the reported result was $\tilde{\Lambda}=89-681$. A reanalysis by the LV collaboration found a new 90\% confidence of $70 \leq\tilde{\Lambda}\leq 720$~\cite{Abbott:2018wiz}. Finally, the LV collaboration reported an additional result, assuming that both merging objects were neutron stars governed by the same EoS~\cite{Abbott:2018exr}. This EoS was based on the Lindblom parametrization~\cite{Lindblom:2010bb} stitched to the SLy EoS for the crust, and resulted in $\tilde{\Lambda}=70-580$ with 90\% confidence. For the different extractions, the lower limit is rather stable, but the upper limit varies from 580-800.





\section{Dense matter equation of state}

Neutron stars are ideal laboratories to test theories of the strong interaction at finite chemical potential and $T=0$. Since neutron stars explore densities from a few gram per cubic centimeter up to 10 times the nuclear saturation density, $n_{\mathrm{sat}}=0.16$~fm~$= 2.7\!\cdot\! 10^{14}$~g~cm$^{-3}$, the knowledge of the EoS is required for densities covering several orders of magnitude. 
While the EoS of the neutron-star crust, reaching up to $n_{\mathrm{sat}}/2$, is rather well constrained, the uncertainty of the EoS increases fast with density and the composition of the inner core of NS is still unknown. Nevertheless, in the density range from $n_{\mathrm{sat}}/2$ up to about $2n_{\mathrm{sat}}$, the neutron-star EoS can be constrained by state-of-the-art nuclear-theory models. The starting point for these constraints are calculations of pure neutron matter (PNM). PNM is an idealized, infinite system consisting solely of neutrons, but it is much easier to compute than systems containing also protons. In contrast to symmetric nuclear matter, PNM is also stable with respect to density fluctuations below $n_{\mathrm{sat}}$, and uniform matter remains the true ground state of PNM at all densities, simplifying its calculation.


In our analysis, we use local chiral effective field theory (EFT) interactions that have been constructed especially for the use in quantum Monte-Carlo (QMC) methods in Refs.~\cite{Lynn:2015jua,Gezerlis:2013ipa,Gezerlis:2014zia,Tews:2015ufa}. These interactions have been successfully tested in light- to medium-mass nuclei and in n-$\alpha$ scattering~\cite{Lynn:2015jua,Lonardoni:2017hgs} and agree with our current knowledge of the empirical parameters of nuclear matter~\cite{Kolomeitsev:2016sjl,Margueron:2017eqc}. In Ref.~\cite{Tews:2018kmu}, these interactions have been used to study neutron matter up to $2n_{\mathrm{sat}}$ with theoretical uncertainty estimates using the AFDMC method.  
For more details on QMC calculations with local chiral interactions we refer the reader to Ref.~\cite{Lynn:2019rdt}. More details on the present approach is given in Ref.~\cite{Tews2019}.


\begin{figure}[t]
\centering
\includegraphics[width=8.0cm]{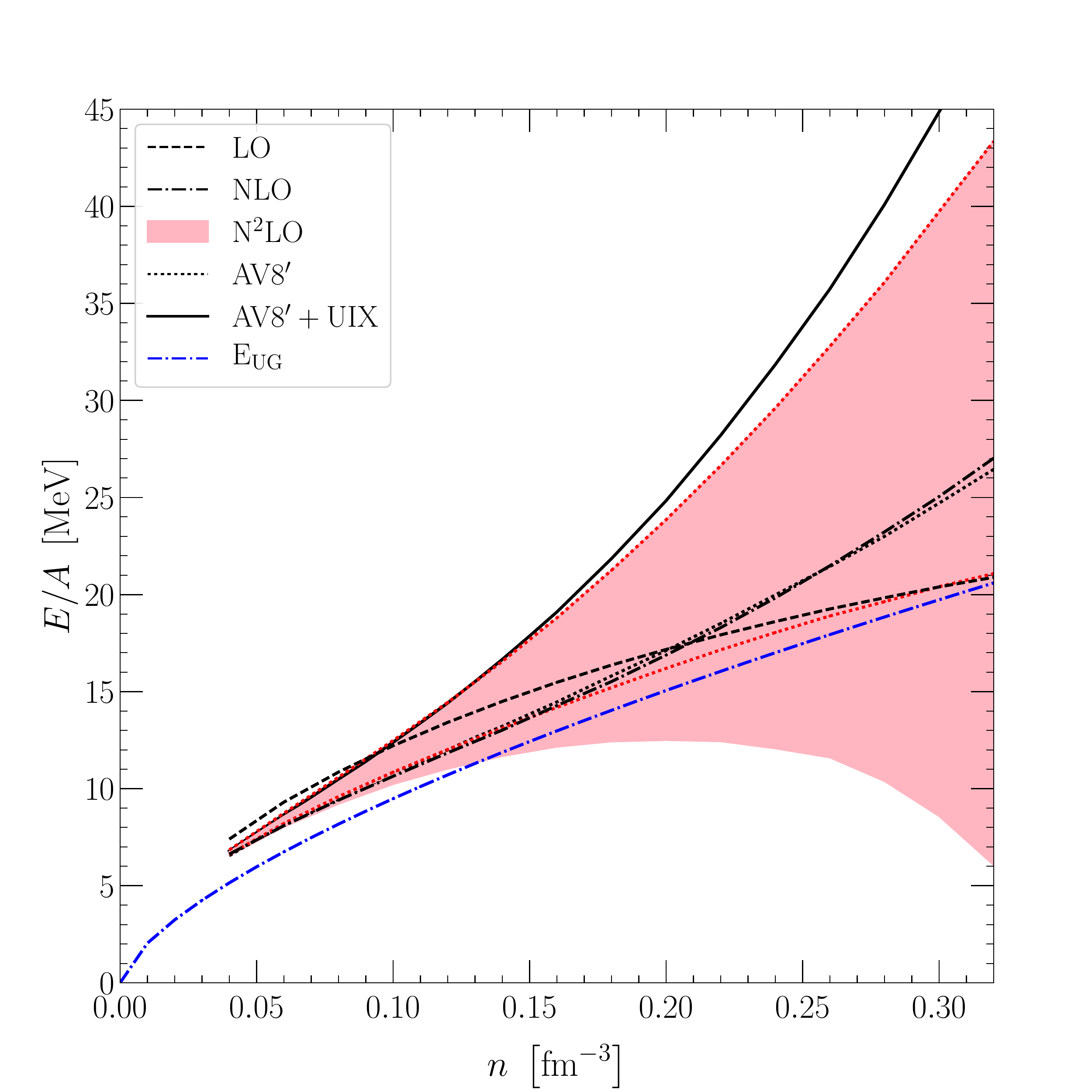}\hspace{0.1cm}
\includegraphics[width=8.0cm]{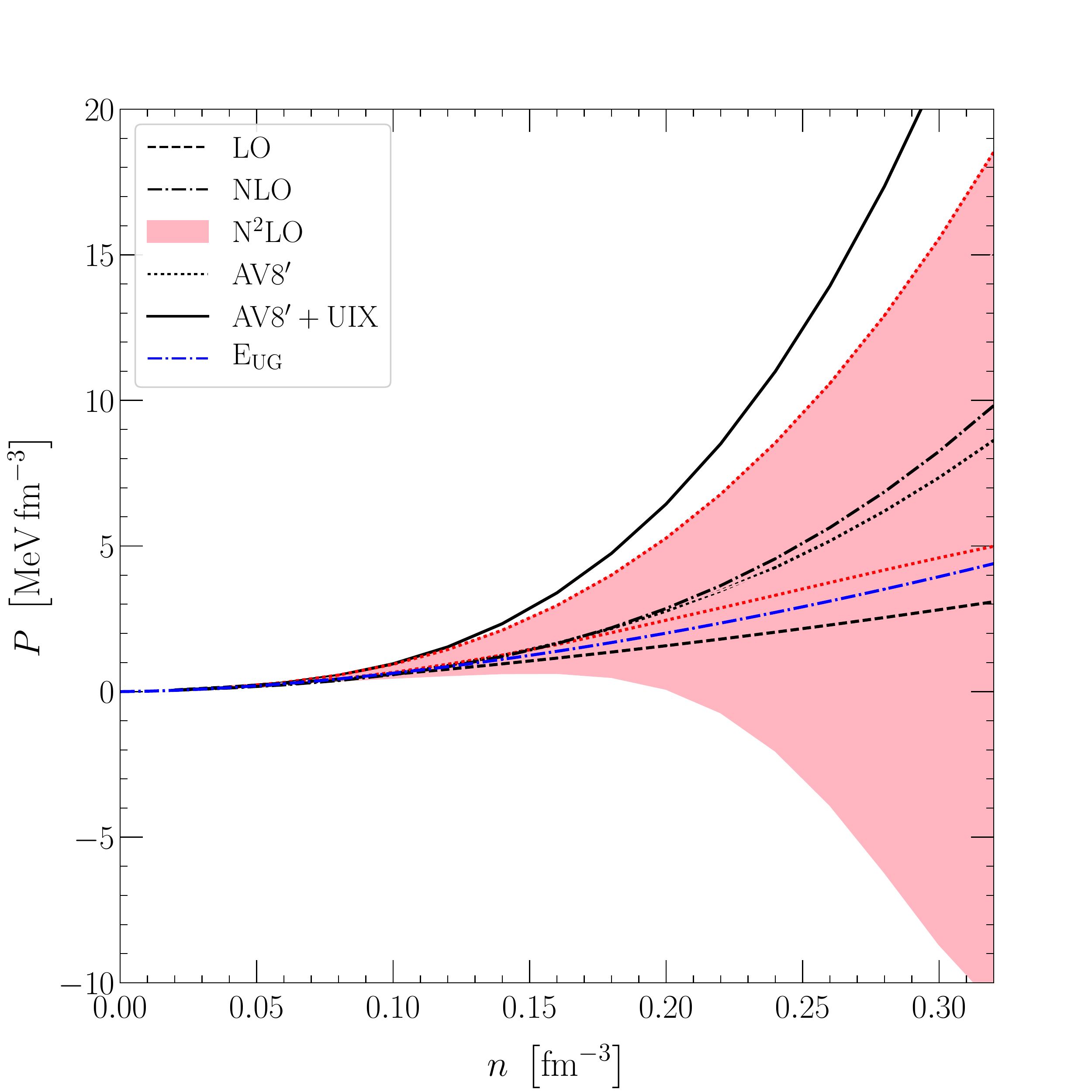}\hspace{0.1cm}
\caption{The energy per particle and pressure of pure neutron matter as functions of baryon density up to $2n_{\rm sat}$. We show the constraints from Ref.~\cite{Tews:2018kmu} based on AFDMC calculations with local chiral potentials at N$^2$LO (red bands). As a comparison, we show results at LO (black dashed lines), NLO (black dashed-dotted lines), as well as calculations using phenomenological $NN$ interactions only (black dotted lines) and including also phenomenological $3N$ forces (black solid lines). We also indicate the unitary-gas bound of Ref.~\cite{Kolomeitsev:2016sjl} (blue dashed-dotted lines) and the part of the uncertainty band that we use for our NS modeling (red dotted lines); see text for more details.}
\label{fig:chiralPNM} 
\end{figure} 

In Fig.~\ref{fig:chiralPNM} we show the results for the energy per particle and pressure of neutron matter at leading order (LO), next-to-leading order (NLO), and at next-to-next-to-leading order (N$^2$LO) with its uncertainty band in a density range going from 0.4~fm$^{-3}$  up to $2n_{\mathrm{sat}}$. We find that the uncertainty bands increase fast with density and are quite sizable at $2 n_{\mathrm{sat}}$. In addition to the results for chiral interactions, we also show in Fig.~\ref{fig:chiralPNM} AFDMC results employing the phenomenological AV8' $NN$ and AV8' $NN$ plus UIX $3N$ interactions as a comparison. It is interesting to note that the AV8' and NLO $NN$ interactions agree very well with each other, which highlights the fact that many-body forces are a considerable source of uncertainty. Finally, we also compare all calculations with the unitary-gas limit of Ref.~\cite{Kolomeitsev:2016sjl}.


In the following, we use this chiral EFT band up to a density $n_{\mathrm{tr}}>n_{\mathrm{sat}}$ to constrain two different modelings for the high density equation of state. By varying $n_{\mathrm{tr}}$ from $n_{\mathrm{sat}}$ to $2n_{\mathrm{sat}}$, we will show that, despite the rapid increase of the uncertainty of the neutron-matter EoS with density, chiral EFT constraints remain extremely useful up to $2 n_{\mathrm{sat}}$.  


\section{Extrapolation schemes for the high density EoS: MM and CSM modelings}

To describe the EoS at higher densities ($n>n_{\mathrm{tr}}$), we will consider two extrapolation schemes rooted in low-density microscopic predictions and widely covering our present uncertainties at higher density. These two schemes are the minimal model or meta-model (MM), based on a smooth extrapolation of chiral EFT results, and the maximal model or speed-of-sound model (CSM), which explores the widest possible domain for the EOS and contains also more drastic behavior with density; see Ref.~\cite{Tews:2018iwm} for the first analysis of GWs with these models and Ref.~\cite{Tews2019} for further analyses. These two models show some overlap for properties of dense neutron-star matter, as suggested from the masquerade phenomenon~\cite{Alford:2004pf}, but also highlight differences: The confrontation of these models with each other and with observations sheds light on the impact of the presence of strong phase transitions at high density, as is detailed hereafter.

The first model that we consider in this analysis, the minimal model or meta-model (MM)~\cite{Margueron:2017eqc,Margueron:2017lup}, assumes the EoS to be smooth enough to be describable in terms of a density expansion about $n_\mathrm{sat}$. 
The MM is described in terms of the empirical parameters of nuclear matter, which are defined as the Taylor coefficients of the density expansion of the energy per particle of symmetric nuclear matter $e_\mathrm{sat}(n)$ and the symmetry energy $s_\mathrm{sym}(n)$, 
\begin{eqnarray}
e_\mathrm{sat}(n) &=& E_\mathrm{sat} + \frac 1 2 K_\mathrm{sat} x^2 + \frac 1 6 Q_\mathrm{sat} x^3 + \frac 1 {24} Z_\mathrm{sat} x^4 + ... \label{eq:esat}\\
s_\mathrm{sym}(n) &=& E_\mathrm{sym} + L_\mathrm{sym} x+ \frac{1}{2} K_\mathrm{sym} x^2 + \frac{1}{6} Q_\mathrm{sym} x^3 +\frac{1}{24} Z_\mathrm{sym} x^4 + ... \,, \label{eq:esym}
\end{eqnarray}
where the expansion parameter $x$ is defined as $x=(n-n_\mathrm{sat})/(3n_\mathrm{sat})$ and $n=n_n+n_p$ is the baryon density, $n_{n/p}$ are the neutron and proton densities.
A good representation of the energy per particle around $n_\mathrm{sat}$ and for small isospin asymmetries $\delta=(n_n-n_p)/n$ can be obtained from the following quadratic approximation,
\begin{equation}
e(n,\delta)=e_\mathrm{sat}(n)+s_\mathrm{sym}(n)\, \delta^2\, .
\end{equation}
The lowest order empirical parameters can be extracted from nuclear experiments~\cite{Margueron:2017eqc}, but typically carry uncertainties. Especially the symmetry-energy parameters are of great interest to the nuclear physics community and considerable effort is invested into a better estimation of their size.

The MM constructs the energy per nucleon as,
\begin{eqnarray}
e^N(n,\delta)=t^{FG*}(n,\delta)+v^N(n,\delta),
\label{eq:MM:energy}
\end{eqnarray}
where the kinetic energy is expressed as 
\begin{eqnarray}
t^{FG^*}(n,\delta) =\frac{t_{sat}^{FG}}{2}\left(\frac{n}{n_{sat}}\right)^{2/3} 
\bigg[ \left( 1+\kappa_{sat}\frac{n}{n_{sat}} \right) f_1(\delta) + \kappa_{sym}\frac{n}{n_{sat}}f_2(\delta)\bigg] ,
\label{eq:MM:kin}
\end{eqnarray}
and the functions $f_1$ and $f_2$ are defined as
\begin{eqnarray}
f_1(\delta) = (1+\delta)^{5/3}+(1-\delta)^{5/3} \, , \;\;\;
f_2(\delta) = \delta \left( (1+\delta)^{5/3}-(1-\delta)^{5/3} \right) .
\end{eqnarray}
The parameters $\kappa_{sat}$ and $\kappa_{sym}$ control the density and asymmetry dependence of the Landau effective mass as ($q$=n or p),
\begin{equation}
\frac{m}{m^*_q(n,\delta)} = 1 + \left( \kappa_{sat} + \tau_3 \kappa_{sym} \delta \right) \frac{n}{n_{sat}} ,
\label{eq:effmass}
\end{equation}
where $\tau_3=1$ for neutrons and -1 for protons.
Taking the limit $\kappa_{sat}=\kappa_{sym}=0$, Eq.~(\ref{eq:MM:kin}) provides the free Fermi gas energy.

The potential energy in Eq.~(\ref{eq:MM:energy}) is expressed as a series expansion in the parameter $x$ and is quadratic in the asymmety parameter $\delta$,
\begin{eqnarray}
v^N(n,\delta)=\sum_{\alpha\geq0}^N \frac{1}{\alpha!}( v_{\alpha}^{sat}+ v_{\alpha}^{sym} \delta^2) x^\alpha u^N_{\alpha}(x) ,
\label{eq:MM:pot}
\end{eqnarray}
where the function $u^N_{\alpha}(x)=1-(-3x)^{N+1-\alpha}\exp(-b n/n_{sat})$ ensures the limit $e^N(n=0,\delta)=0$.
The parameter $b$ is taken large enough for the function $u^N_{\alpha}$ to fall sufficiently fast with density and to not contribute at densities above $n_{sat}$. A typical value is $b=10\ln2\approx 6.93$ such that the exponential function is $1/2$ for $n=n_{sat}/10$.
The MM parameters $v_{\alpha}^{sat}$ and $v_{\alpha}^{sym}$ are simply expressed in terms of the empirical parameters~\cite{Margueron:2017eqc}. 
To obtain the neutron-star EoS, we extend our models to $\beta$-equilibrium and include a crust as described in Ref.~\cite{Margueron:2017lup}. By varying the empirical parameters within their known or estimated uncertainties, it was shown that the MM can reproduce many existing neutron-star EoS that are based on the assumption that a nuclear description is valid at all densities probed in neutron stars. Therefore, this model is a reliable representation for EoS without exotic phases of matter separated from the nucleonic phase through strong phase transitions.

\begin{table}[t]
\centering
\setlength{\tabcolsep}{8pt}
\renewcommand{\arraystretch}{1.2}
\begin{tabular}{cccccccccccc}
\hline
$P_{\alpha}$ & $E_{sat}$ & $E_{sym}$ & $n_{sat}$ & $L_{sym}$ & $K_{sat}$ & $K_{sym}$ & $Q_{sat}$ & $Q_{sym}$ & $Z_{sat}$ & $Z_{sym}$ & $b$\\
                    & MeV           & MeV           & fm$^{-3}$           & MeV           & MeV           & MeV           & MeV           & MeV           & MeV           & MeV & \\
\hline
Max             & -15 & 38 & 0.17 & 90  &  270  &  200  &   1000 &   2000 &   3000 &    3000 & 14 \\
Min              & -17 & 26 & 0.15 & 20  &  190  & -400  &  -1000 &   -2000 &  -3000 & -3000 &1 \\
\hline
\end{tabular}
\caption{Empirical parameters and their domain of variation entering into the definition of the MM~(\ref{eq:MM:energy}). The parameters $\kappa_{sat}$ and $\kappa_{sym}$  are fixed such that $m_{sat}^*/m=0.75$ in symmetric matter and $m_n^*/m-m_p^*/m=-0.1$ in neutron matter.}
\label{tab:epbound}
\end{table}

In the following, the parameter space for the MM will be explored within a Markov-Chain Monte-Carlo algorithm, where the MM parameters are allowed to freely evolve inside the boundaries given in Table.~\ref{tab:epbound}. The resulting models satisfy the chiral EFT predictions in neutron matter for the energy per particle and the pressure up to $n_\mathrm{tr}$, causality, stability, positiveness of the symmetry energy ($s_\mathrm{sym}(n)>0$), and also reach the maximum observed neutron-star mass $M_\mathrm{max}^\mathrm{obs}$. The maximum density associated to each EoS within the MM is given either by the break-down of causality, stability, or positiveness of the symmetry energy condition, or by the end point of the stable neutron-star branch.


The second model that we consider in this analysis, the maximal model (CSM), is based on an extension of the speed of sound in neutron-star matter. Starting from the pure neutron matter calculations, we construct the neutron-star EoS up to $n_\mathrm{tr}$ by constructing a crust as described in Ref.~\cite{Tews:2016ofv} and extending the neutron-matter results to $\beta$ equilibrium above the crust-core transition. Having constructed the EoS up to $n_\mathrm{tr}$ we compute the speed of sound, 
\begin{equation}
c_S^2 = \frac{\partial p(\epsilon)}{\partial \epsilon}\,,
\end{equation}
where $p$ is the pressure and $\epsilon$ is the energy density. Above $n_\mathrm{tr}$, we parametrize the speed of sound in a very general way: we randomly sample a set of points $c_S^2(n)$, where the values for $c_S$ have to be positive and are limited by the speed of light (stability and causality), and interpolate between the different sampling points using linear segments. The individual points are randomly distributed in the interval $n_\mathrm{tr}-12 n_\mathrm{sat}$.  From the resulting speed-of-sound curve, we reconstruct the EoS step-by-step starting at $n_\mathrm{tr}$, where $\epsilon(n_\mathrm{tr})$,  
$p(n_\mathrm{tr})$, and $\epsilon'(n_\mathrm{tr})$ are known:
\begin{eqnarray}
n_{i+1}= n_i + \Delta n, \;\;\;
\epsilon_{i+1} = \epsilon_i +\Delta\epsilon= \epsilon_i + \Delta n \cdot \left(\frac{\epsilon_i+p_i}{n_i}\right), \;\;\;
p_{i+1} = p_i + c_S^2 (n_i) \cdot \Delta \epsilon\,,
\end{eqnarray}
where $i=0$ defines the transition density $n_\mathrm{tr}$. In the second line we have used the thermodynamic relation $p=n \partial \epsilon/\partial n -\epsilon$, which is valid at zero temperature. 
In that way, we iteratively obtain the high-density EoS. We have explored extensions for a varying number of $c_S^2(n)$ points, i.e., for 5-10 points, and found that the differences between these extensions are marginal. We, therefore, choose 6 sampling points. For each sampled EoS, we generate a second version which includes a strong first-order phase transition with a random onset density and width, to explicitly explore such extreme density behavior.

The CSM for neutron-star applications was introduced in Ref.~\cite{Tews:2018kmu}, and represents and extension of the model of Ref.~\cite{Alford:2013aca}. A similar model was used in Ref.~\cite{Greif:2018njt}. However, in contrast to Ref.~\cite{Tews:2018kmu} we have extended this model to explore the complete allowed parameter space for the speed of sound, by abandoning the specific functional form of Ref.~\cite{Tews:2018kmu} in favor of an extension using linear segments. This more conservative choice leads to slightly larger uncertainty bands, but allows us to make more definitive statements about neutron-star properties. The resulting EoS parameterizations represent possible neutron-star EoS and may include drastic density dependences, e.g., strong phase transitions which lead to intervals with a drastic softening or stiffening of the EoS. 
This represents a stark contrast to the MM, which does not include such behavior, and might give insights into the constituents of neutron-star matter at high-densities. The predictions of the CSM represent the widest possible domain for the respective neutron-star observables consistent with the low density input from chiral EFT. If observations outside of this domain were to be made, this would imply a breakdown of nuclear EFTs at densities below the corresponding $n_\mathrm{tr}$. 

Since the CSM represents very general EoSs only governed by the density dependence of the speed-of-sound, it does not allow any statements about possible degrees of freedom. In this sense, it is very similar to extensions using piecewise polytropes which were introduced in Ref.~\cite{Read:2008iy} and have been used extensively to determine neutron-star properties; see, e.g., Ref.~\cite{Hebeler:2013nza,Raithel:2016bux,Annala:2017llu}. However, in contrast to polytropic extensions, in the CSM the speed of sound is continuous except when first-order phase transition are explicitly accounted for. This is important for the study of the tidal polarizabilities, where $c_S^{-1}$ enters.


\begin{figure}[t]
\centering
\includegraphics[trim= 0.0cm 0 0 0, clip=,width=0.33\columnwidth]{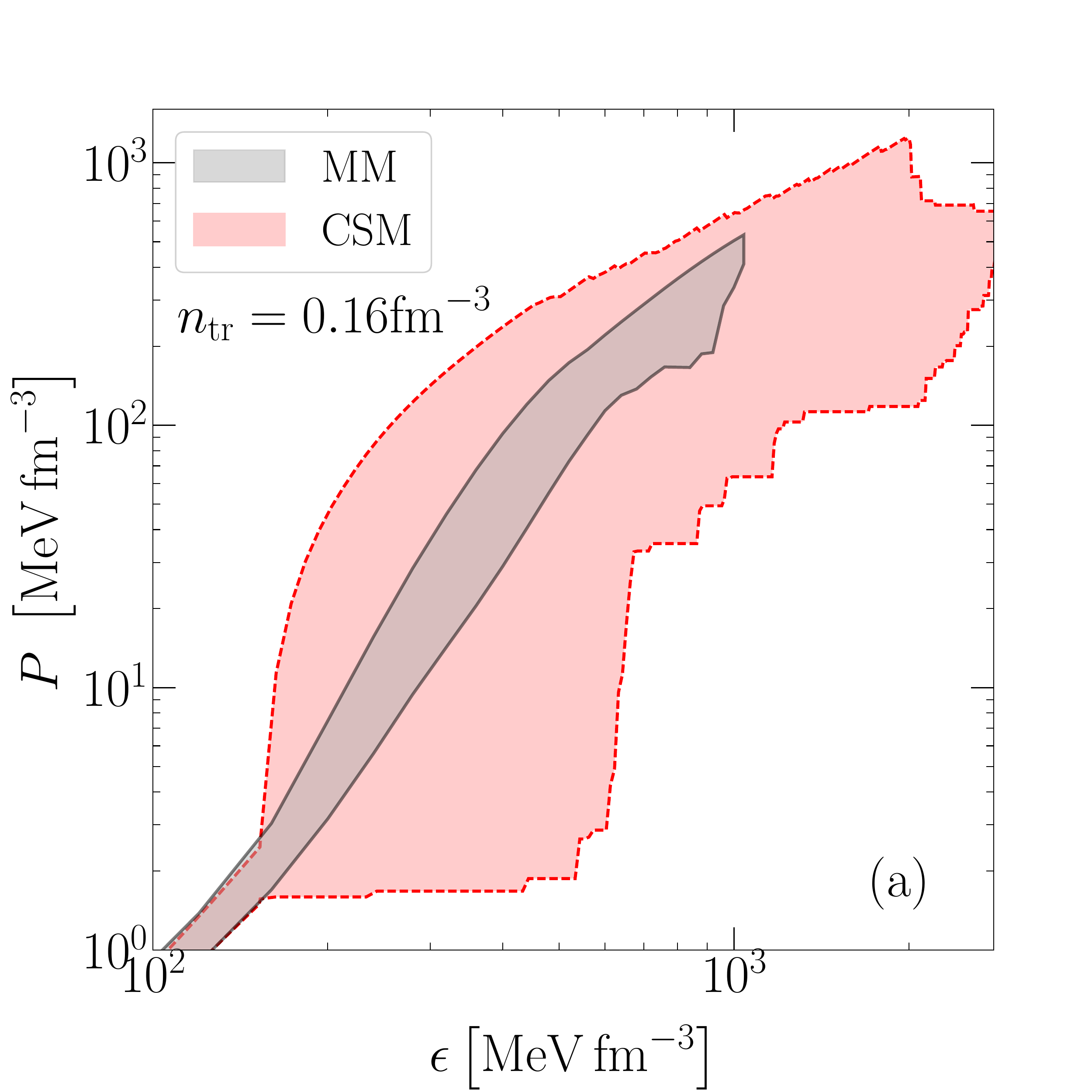}
\includegraphics[trim= 0.0cm 0 0 0, clip=,width=0.33\columnwidth]{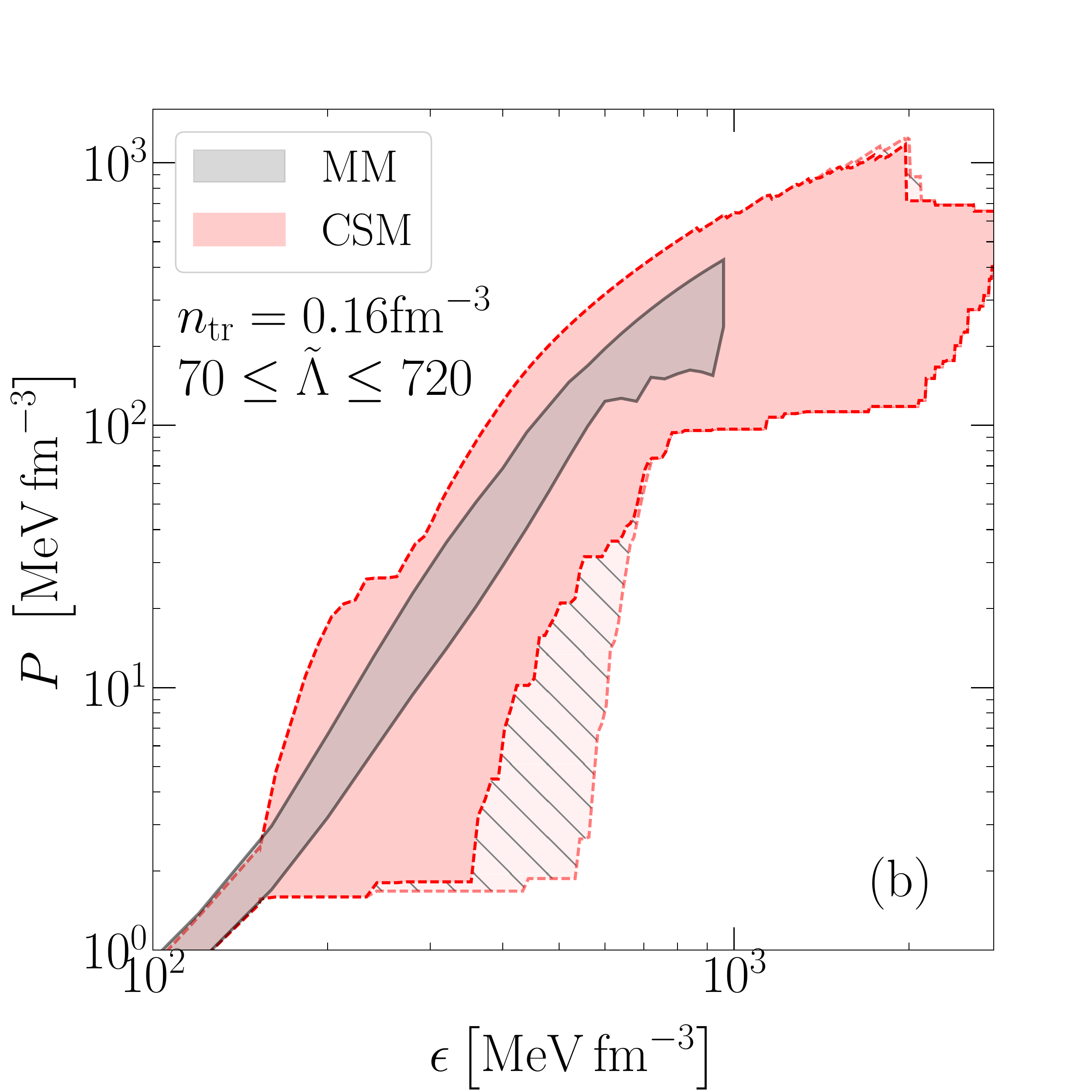}
\includegraphics[trim= 0.0cm 0 0 0, clip=,width=0.33\columnwidth]{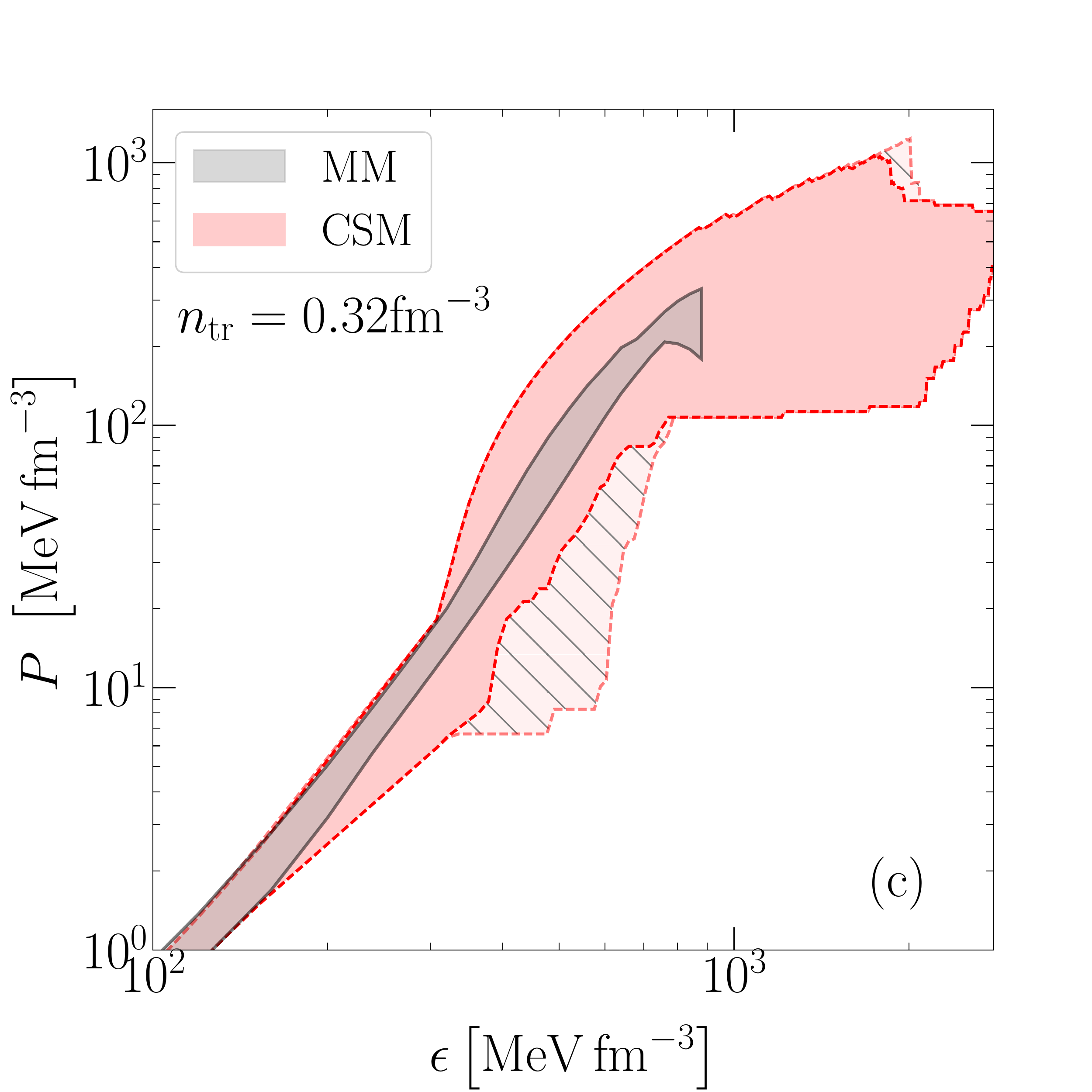}
\caption{\label{fig:EpsPcomp}
Comparison of the allowed EoS envelopes for the MM (black bands) and the CSM (red bands). We show three cases: a) the most general case, where $n_\mathrm{tr}=n_\mathrm{sat}$ and only $M_\mathrm{max}\geq 1.9 M_{\odot}$ is enforced, b) for $n_\mathrm{tr}=n_\mathrm{sat}$ when enforcing $70\leq \tilde{\Lambda} \leq 720$ and c) for $n_\mathrm{tr}=2 n_\mathrm{sat}$. When additionally enforcing $R_{1.6}\geq 10.68$ km, the hatched regions are excluded.
}
\end{figure}

For both the MM and CSM we generate thousands of EoSs that are consistent with low-density constraints from chiral EFT. In addition, the observations of heavy two-solar-mass pulsars in recent years~\cite{Demorest2010,Antoniadis2013,Fonseca2016} place important additional constraints on these EoSs, which we enforce by requiring  $M_\mathrm{max}>M_\mathrm{max}^\mathrm{obs}$ for all our EoSs. 
To be conservative, as the limit for $M_\mathrm{max}^\mathrm{obs}$ we choose the centroid of the maximum observed mass minus twice the error-bar on the observation. For the two heaviest neutron stars observed up to now~\cite{Demorest2010,Antoniadis2013,Fonseca2016}, this gives $M_\mathrm{max}^\mathrm{obs}\approx 1.9 M_\odot$. 

\begin{figure}[t]
\centering
\includegraphics[trim= 0.0cm 0 0 0, clip=,width=0.33\columnwidth]{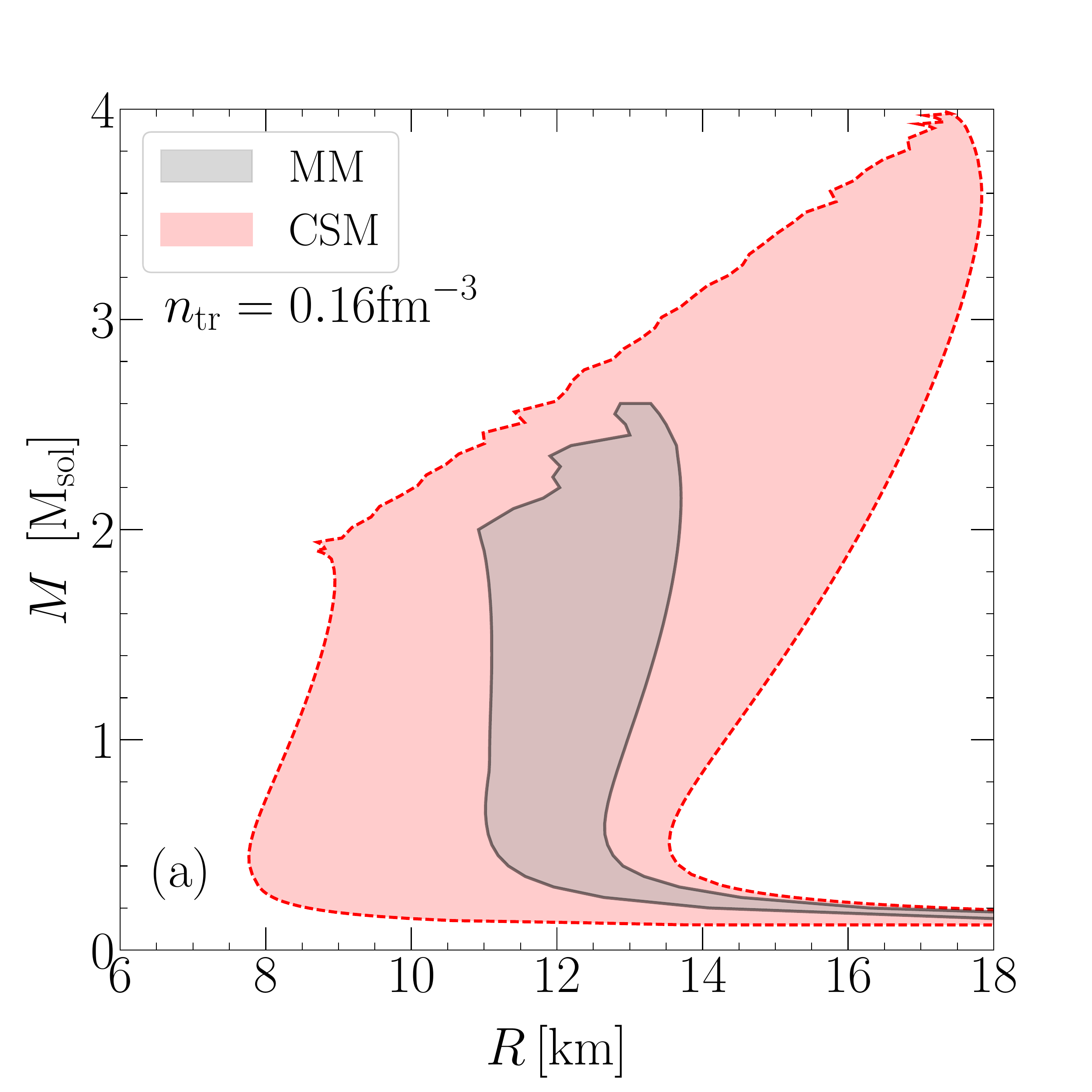}
\includegraphics[trim= 0.0cm 0 0 0, clip=,width=0.33\columnwidth]{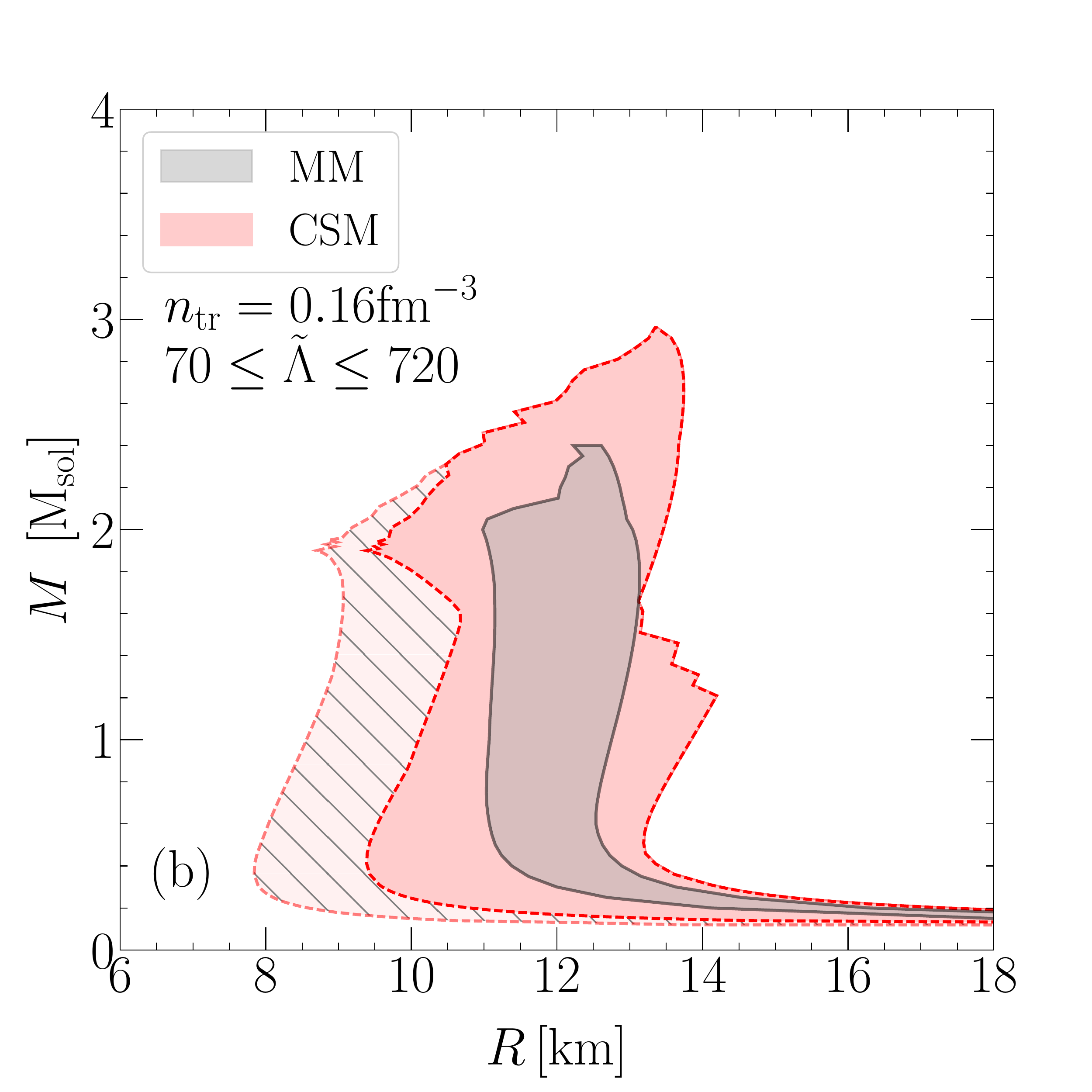}
\includegraphics[trim= 0.0cm 0 0 0, clip=,width=0.33\columnwidth]{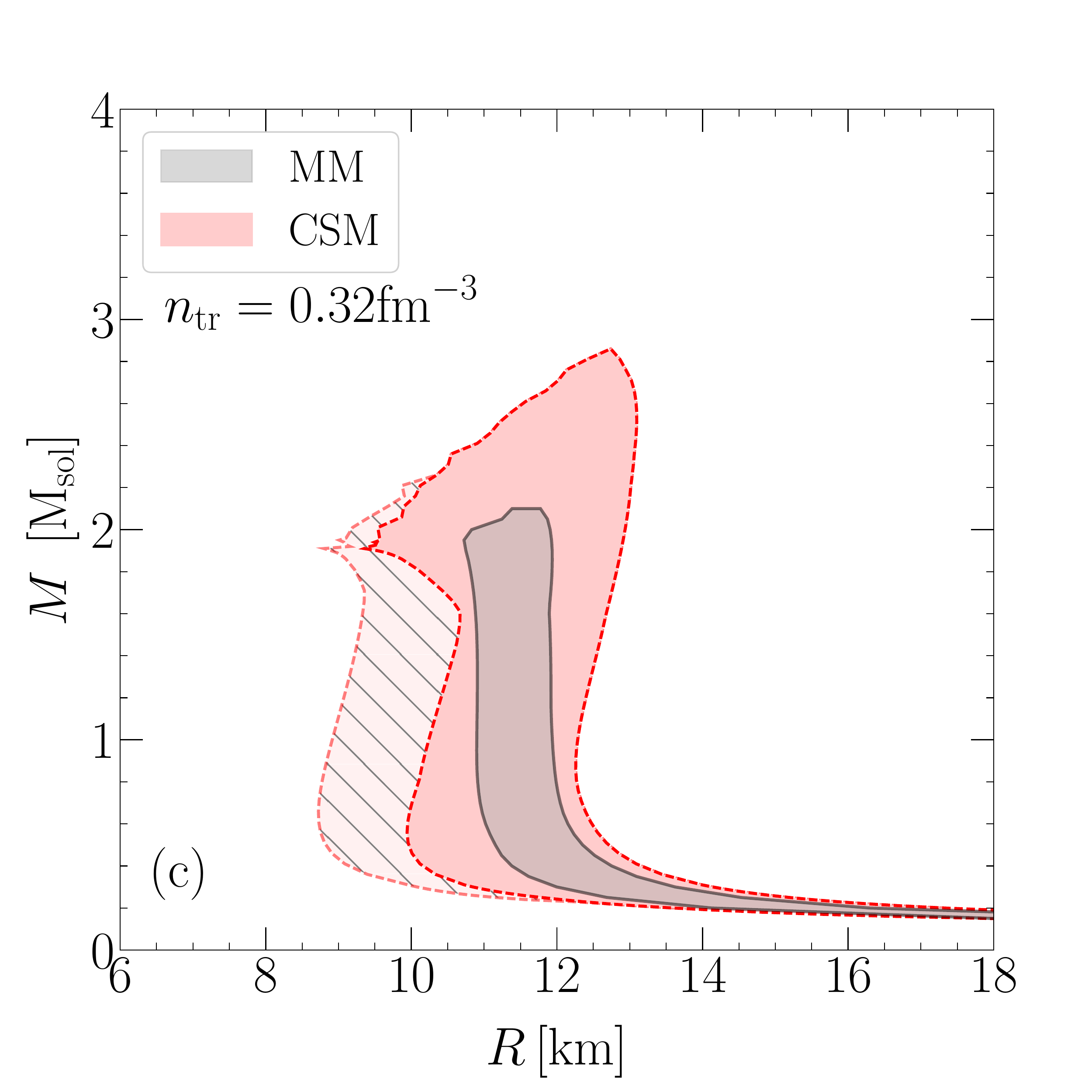}
\caption{\label{fig:MRcomp}
Comparison of the allowed MR envelopes for the MM (black bands) and the CSM (red bands). We show three cases: a) the most general case, where $n_{\mathrm{tr}}=n_{\mathrm{sat}}$ and only $M_{\rm{max}}\geq 1.9 M_{\odot}$ is enforced, b) for $n_\mathrm{tr}=n_\mathrm{sat}$ when enforcing $70\leq \tilde{\Lambda} \leq 720$, and c) for $n_\mathrm{tr}=2 n_\mathrm{sat}$. When additionally enforcing $R_{1.6}\geq 10.68$ km, the hatched regions are excluded.
}
\end{figure}

We now compare the predictions of both the MM (black bands with solid contour) and CSM (red bands with dotted contour) for the EoS of neutron-star matter, see Fig.~\ref{fig:EpsPcomp}, and the mass-radius (MR) relation, see Fig.~\ref{fig:MRcomp}. In the respective figures, we show the EoS and MR envelopes for $n_\mathrm{tr}=n_\mathrm{sat}$ [panels (a)] and for $n_\mathrm{tr}=2 n_\mathrm{sat}$ [panels (c)]. In all cases, the MM is a subset of the CSM, as expected. Also, the two models, which treat the neutron-star crust with different prescriptions, show excellent agreement at low densities. For $n_\mathrm{tr}=n_\mathrm{sat}$, the MM and CSM EoSs agree very well up to  $n_\mathrm{tr}$, while for $n_\mathrm{tr}=2 n_\mathrm{sat}$ the MM only samples a subset of the chiral EFT input, because the $M_\mathrm{max}^\mathrm{obs}$ constraint forces the EoS to be sufficiently stiff which excludes the softest low-density neutron-matter EoS. This is a consequence of the smooth density expansion around $n_\mathrm{sat}$ in the MM. In the CSM, instead, a non-smooth stiffening of these softest EoS at higher densities can help stabilize heavy neutron stars, which is why the complete low-density band from chiral EFT is sampled.  We also find that going from $n_\mathrm{tr}=n_\mathrm{sat}$ to $n_\mathrm{tr}=2 n_\mathrm{sat}$ allows to considerable reduce the EoS uncertainty for the CSM. The MM uncertainty is also slightly reduced and the MM band gets narrower. These results show that even though the theoretical uncertainties in the neutron-matter EoS increase fast in the density range $1-2 n_\mathrm{sat}$, the additional information provided allows to substantially reduce uncertainties in the CSM EoS:   essentially, the chiral EFT constraint excludes the possibility of phase transitions in the region going from 1 to $2n_{sat}$. The impact of phase transitions above $2n_\mathrm{sat}$ on the EoS is very much reduced compared to the case where they are allowed to appear at lower densities, because we impose the $M_\mathrm{max}^\mathrm{obs}$ constraint. A large domain of soft CSM EoSs is, thus, excluded. The stiff MM and CSM EoS are very close up to $2n_\mathrm{sat}$, as expected.

These observations are also reflected in the MR predictions of both models shown in Fig.~\ref{fig:MRcomp}. For $n_\mathrm{tr}=n_\mathrm{sat}$ [panel (a)], the CSM (MM) leads to a radius range of a typical neutron star of $1.4 M_{\odot}$ of $8.4-15.2$ km ($10.9-13.5$ km). This range reduces dramatically for $n_\mathrm{tr}=2 n_\mathrm{sat}$ [panel (c)], where we find $8.7-12.6$ km ($10.9-12.0$ km) for the CSM (MM). 
In the last case, the radius uncertainty for a typical neutron star is only about 1 km in the MM, compatible with the expected uncertainty of the NICER mission~\cite{NICER1}. This allows for a possible tension between the MM and NICER predictions. If such an observation should be made in the near future, we will be able to better constrain dense-matter phase transitions. In contrast, the CSM, which includes EoS with sudden softening or stiffening at higher densities, dramatically extends the allowed envelopes for the EoS and the MR relation as compared with the MM. These differences in the predictions of the MM and CSM can be used to identify regions for the neutron-star observables, for which statements about the constituents of matter might be possible. For example, the observation of a typical neutron star with a radius of 10 km would imply the existence of a softening phase transition, that would hint on new phases of matter appearing in the core of neutron stars. Instead, in regions were both the MM and CSM agree, the masquerade problem does not allow statements about the constituents of neutron-star matter at high densities~\cite{Alford:2004pf}.


It is interesting to look at areas of constant $\Lambda$ within the MR plane. In this case, the relation of neutron-star mass and radius is given by 
\begin{eqnarray}
M&=\left(\frac32 \frac{\Lambda}{k_2}\right)^{-\frac15} \frac{c^2}{G} R\,,
\end{eqnarray}
leading to the following scaling relation,
\begin{eqnarray}
\left(\frac{M}{M_{\odot}}\right)&=0.6243 \left(\frac{\Lambda}{k_2}\right)^{-\frac15} \left(\frac{R}{1 \km}\right)\,.
\label{eq:scaling}
\end{eqnarray}

\begin{figure}[t]
\centering
\includegraphics[trim= 0.0cm 0 0 0, clip=,width=0.5\columnwidth]{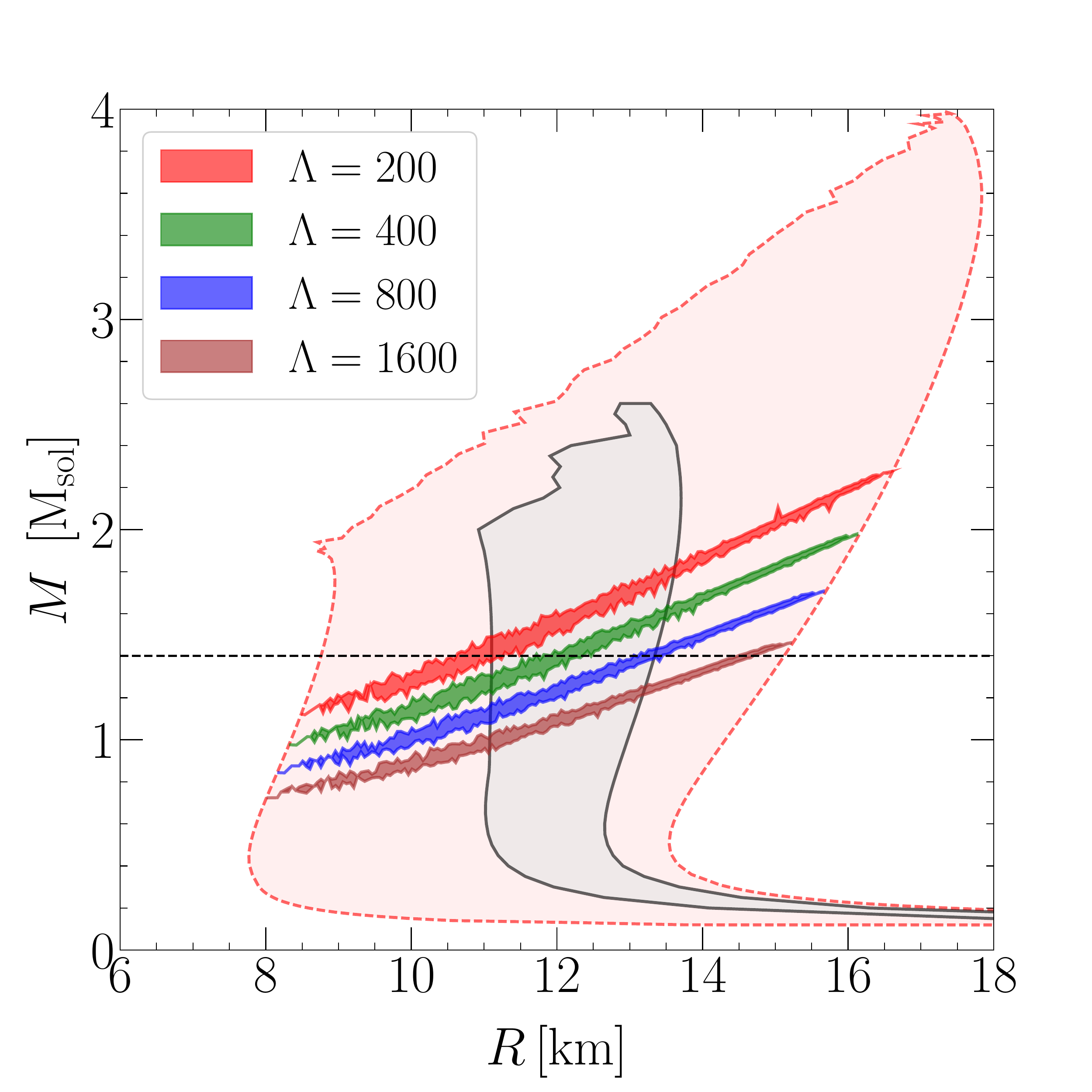}
\caption{\label{fig:MRLam}
Mass-radius envelopes for $n_{\rm tr}=n_{\rm sat}$ of Fig.~\ref{fig:MRcomp}(a) and areas of constant $\Lambda$ for all CSM EOS parametrizations. We show areas for $\Lambda=200$ (red), $\Lambda=400$ (green), $\Lambda=800$ (blue), and for $\Lambda=1600$ (brown). For a typical $1.4 M_{\odot}$ neutron star (horizontal dashed line), a constraint on $\Lambda$ is equivalent to a radius constraint. The corresponding values for the MM (not shown) always lie withing the areas for the CSM. 
}
\end{figure} 

For constant $\Lambda$, this implies an almost linear relationship between M and R, because the love number $k_2$ does not vary strongly in that case. In addition, for different values of $\Lambda$, the slopes are rather similar due to the small exponent $-1/5$. In Fig.~\ref{fig:MRLam}, we plot the mass-radius relation for $n_\mathrm{tr}=n_\mathrm{sat}$ for the CSM, together with areas of constant $\Lambda$. In particular, we show areas for $\Lambda=200, 400, 800$, and $1600$.

While there is a tight correlation between radii and tidal polarizabilities, from Fig.~\ref{fig:MRLam} one can see that both quantities still provide complementary information. For example, an exact observation of the tidal polarizability of a neutron star, i.e., with vanishing uncertainty, would still lead to a remaining uncertainty for the radius of a typical $1.4 M_{\odot}$ neutron star. To be specific, for $\Lambda=200$, the remaining radius uncertainty is still $\approx 1$ km, compatible with the expected uncertainty of NICER~\cite{NICER1}. For larger values of $\Lambda$ this uncertainty decreases and for $\Lambda=800$ it is only $\approx 0.5$ km. However, based on GW170817 values larger than $720$ are ruled out for typical neutron stars. Hence, both tidal deformabilities and radii offer complementary information on neutron-star global structure. 

Finally, from Eq.~(\ref{eq:scaling}), one can infer the following fit, with $a=0.406435$ and $b= 68.5$,
\begin{eqnarray}
\left(\frac{M}{M_{\odot}}\right)&= \frac{a}{(b+\Lambda)^{1/5}} \left(\frac{R}{1 \km}\right)\,.
\label{eq:scalingFit}
\end{eqnarray}

\subsection{Impact of varying $n_\mathrm{tr}$ and the validity of chiral EFT predictions}

\begin{figure}[t]
\centering
\includegraphics[trim= 0.0cm 0 0 0, clip=,width=0.5\columnwidth]{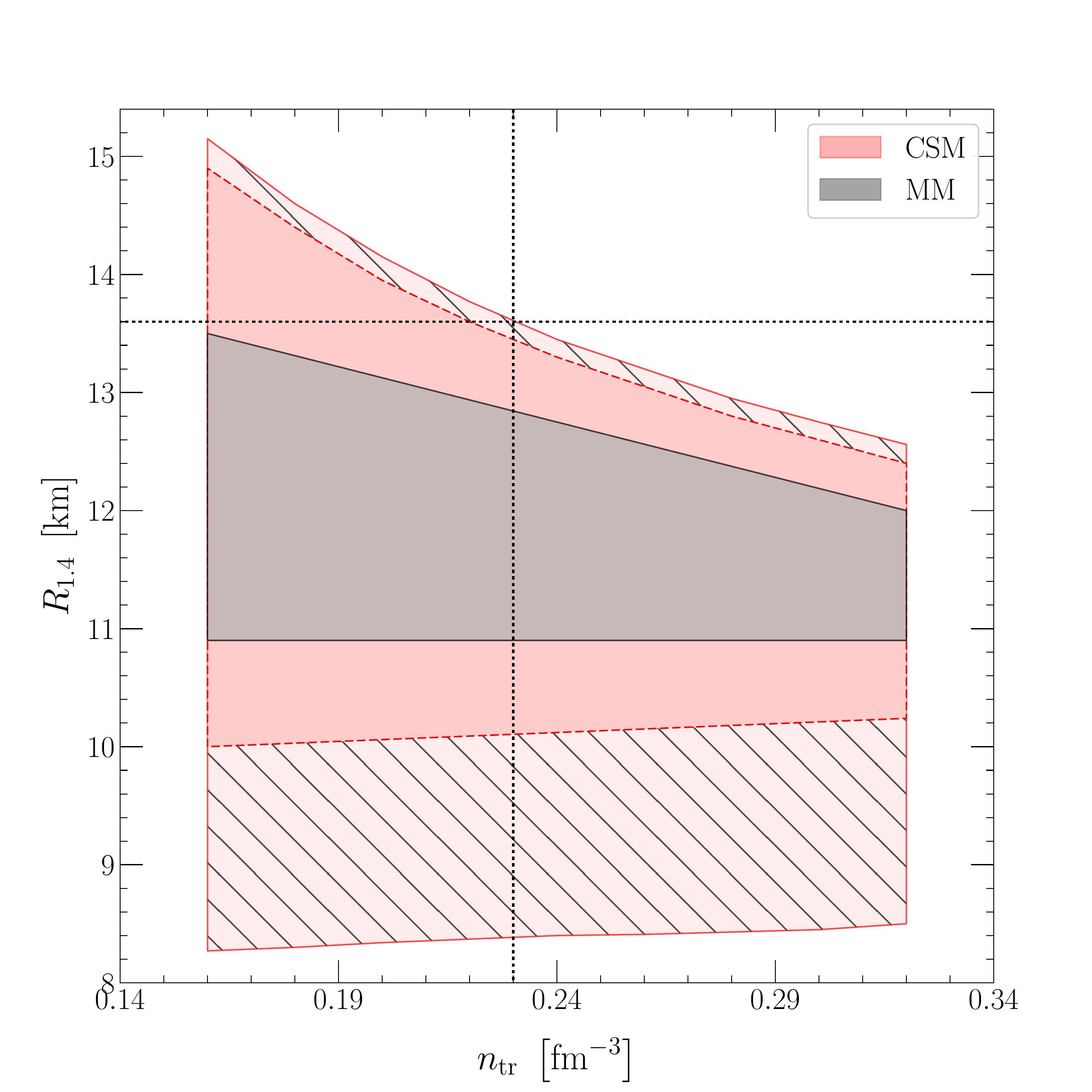}
\includegraphics[trim= 0.0cm 0 0 0, clip=,width=0.5\columnwidth]{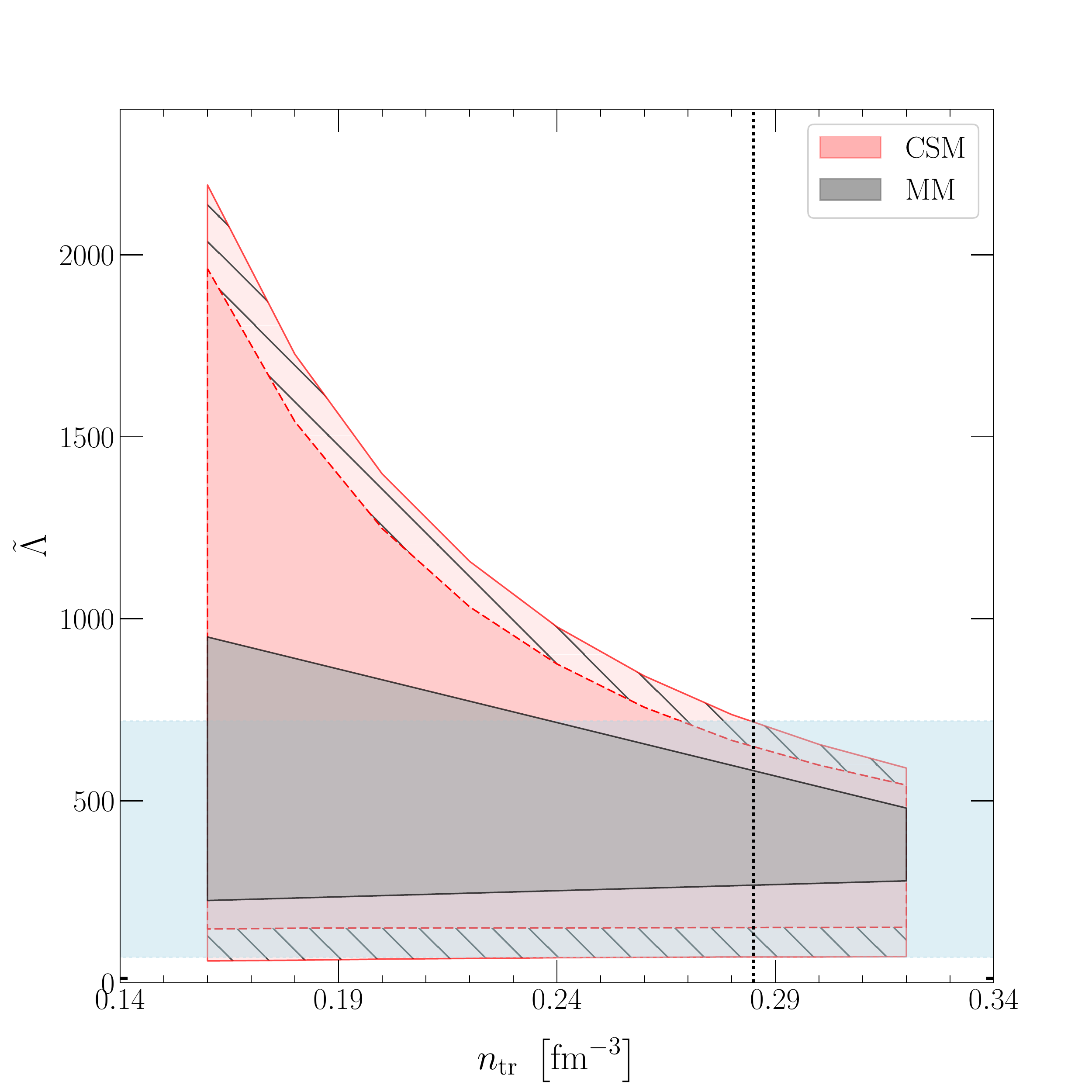}
\caption{\label{fig:ntrRminRmax}
Radius of a typical $1.4 M_{\odot}$ neutron star, $R_{1.4}$ (left), and $\tilde{\Lambda}$ for $M_{\rm chirp}=1.186 M_{\odot}$ (right) as functions of $n_{\rm{tr}}$. We show the envelopes for the CSM in red and for the MM in black. For the CSM, when requiring $c_S^2\leq 0.5$ instead of $c_S^2\leq1.0$, the hatched areas are excluded. We also indicate the constraints from GW170817 and the values of $n_{\rm{tr}}$, above which nuclear-theory input alone becomes more constraining than observations.}
\end{figure} 

These present studies as well as the one of Refs.~\cite{Tews:2018iwm,Lim:2018bkq,Tews2019} are the first to use chiral EFT calculations of the neutron matter EoS up to twice nuclear saturation density with reliable error estimates~\cite{Tews:2018kmu} to compute tidal polarizabilities for GW170817. Reliable uncertainty estimates are critical for understanding the impact that GW detections will have on elucidating the properties of dense matter inside neutron stars, and theoretical calculations of the dense-matter EoS without uncertainty estimates are of limited value for a meaningful analysis of GW data. Uncertainty estimates have shown that chiral EFT input remains useful up to $2 n_\mathrm{sat}$, and we find, in contrast to other recent publications~\cite{Annala:2017llu,Fattoyev:2017jql,Most:2018hfd} which had limited the chiral EFT input up to only $n_\mathrm{sat}$, that GW170817 does \emph{not} provide new insight about the EoS that cannot be obtained from current nuclear physics knowledge. This message tempers claims made in these recent publications which state that the upper limit on the tidal polarizability derived from GW data rules out stiff nuclear EoS. While this inference is correct, such stiff EoSs are already ruled out based on state-of-the-art nuclear Hamiltonians. In other words, models of dense matter excluded by the upper limit on the tidal deformability from GW170817 are already incompatible with the current microscopic EoSs at densities where error estimates can still be justified. 

Nevertheless, the reliability of chiral interactions at these densities has been questioned. Although the convergence of the chiral expansion cannot be strictly proven in this density range, we present arguments to show that the order-by-order convergence of the chiral expansion for the EoS up to $2n_\mathrm{sat}$ is still reasonable. First, the breakdown of the chiral expansion is not easy to define in terms of an upper value for the density. As an illustration, at saturation density the expansion parameter is less than 0.6 and it increases by only about 25\% over the density interval $1-2 n_\mathrm{sat}$. So the expansion parameter is not dramatically worst at $2 n_\mathrm{sat}$ compared to $n_\mathrm{sat}$. Second, Ref.~\cite{Tews:2018kmu} analyzed the order-by-order convergence of the employed Hamiltonians at $2 n_\mathrm{sat}$, and showed that, even though the reliability naturally decreases with increasing density, the order-by-order convergence remains reasonable and consistent with simple power counting arguments within the theoretical uncertainty estimates. Nevertheless, densities around $2 n_\mathrm{sat}$ seem to provide an upper limit to the applicability of the chiral Hamiltonians we use in this work.

To support our main statement - namely that the constraints from GW170817 are compatible with but less restrictive than predictions of the EoS based on realistic nuclear potentials and do not yield specific new information about nuclear Hamiltonians or about possible phase transitions at supra-nuclear density - in this context, we investigate which density range for chiral EFT input is sufficient to justify our statement. We present the total uncertainty ranges for $R_{1.4}$ (left panel) and $\tilde{\Lambda}$ for $M_\mathrm{chirp}=1.186 M_{\odot}$(right panel) as functions of the density $n_\mathrm{tr}$ in Fig.~\ref{fig:ntrRminRmax}. For $R_{1.4}$, we indicate the upper limit on the radii of Ref.~\cite{Annala:2017llu}, $R_{1.4}\leq 13.6$ km, which was obtained using $n_\mathrm{tr}=n_\mathrm{sat}$ and the LV constraint (horizontal dotted line). We find that the CSM alone constrains the radii to be smaller than this bound for $n_\mathrm{tr}>0.23 \fmiq \approx 1.44 n_\mathrm{sat}$ (an 11\% increase of the expansion parameter compared to $n_\mathrm{sat}$). For the tidal polarizability, we indicate the LV constraint as a horizontal blue band and find that the CSM leads to $\tilde{\Lambda}\leq 720$ as soon as $n_\mathrm{tr}> 0.285 \fmiq \approx 1.78 n_\mathrm{sat}$ (a 20\% increase of the expansion parameter compared to $n_\mathrm{sat}$). We would like to emphasize that these crucial values for $n_\mathrm{tr}$ for both observables do not necessarily have to agree, as seen in Fig.~\ref{fig:ntrRminRmax}. The reason is that the upper limit on $\tilde{\Lambda}$ depends on $q$ while $R_{1.4}$ does not. 
Chiral EFT input becomes compatible with this value at $n_\mathrm{tr}\sim 0.23 \fmiq$, in agreement with the value for $R_{1.4}$. At these values for $n_\mathrm{tr}$, in particular at $1.44 n_\mathrm{sat}$, arguments for the validity of chiral interactions remain even stronger,  which strengthens the validity of our main statement.

Finally, the value of $n_\mathrm{tr}$ also affects the speed of sound inside neutron stars. The speed of sound is expected to approach the conformal limit of $c_S^2=1/3$ at very high densities~\cite{Kurkela:2010}. In neutron stars, though, it is not clear if this conformal limit is useful or not. As discussed in detail in Ref.~\cite{Tews:2018kmu}, the neutron-matter EoS up to $n_\mathrm{tr}=2 n_\mathrm{sat}$ requires the speed of sound to pass the conformal limit to be sufficiently stiff to stabilize the observed two-solar-mass neutron stars. In fact, for chiral models the speed of sound has to increase beyond the conformal limit for $n_\mathrm{tr}>0.28 \fmiq$ and even for phenomenological nuclear Hamiltonians, which lead to stiffer neutron-matter EoS, this statement remains valid for $n_\mathrm{tr}>0.31 \fmiq$. While there might be EoS that are much stiffer below $2 n_\mathrm{sat}$ and, hence, stabilize the heaviest neutron stars while still obeying the conformal limit, such EoS are ruled out by modern nuclear Hamiltonians. 
Therefore, the neutron-matter EoS up to $2 n_\mathrm{sat}$ for state-of-the-art nuclear Hamiltonians requires the speed of sound in neutron stars to experience a non-monotonous behavior, i.e, increasing beyond $c_S^2=1/3$ but decreasing at higher densities to approach this limit.
For example, for chiral EFT interactions and $n_\mathrm{tr}=2 n_\mathrm{sat}$, the speed of sound has to reach values $c_S^2\geq 0.4$. The question remains, though, which forms of strongly-interacting matter lead to such a behavior for the speed of sound. 
The hatched areas in Fig.~\ref{fig:ntrRminRmax} represent the predictions based on EoS for which $c_S^2\geq 0.5$.
Excluding these EoS
slightly reduces the upper bound on neutron-star radii but it would mostly rule out low-radius neutron stars. The reason is that neutron stars can have very low radii only for strong first-order phase transitions with small onset densities. To simultaneously support $2M_{\odot}$ neutron stars, the EOSs has to experience a sudden subsequent stiffening, i.e., the speed of sound has to increase dramatically. For a larger possible speed of sound, stronger phase transitions are allowed, which leads to stars with small radii. Limits on $c_S^2$, on the other hand, rule out the strongest phase transition, and increase the smallest possible radius. For $c_S^2\leq 0.5$, the lower limit on the radius of a $1.4M_{\odot}$ neutron star is of the order of 10 km, of the order of the constraint of Ref.~\cite{Bauswein:2017vtn}.

In the next years additional neutron-star merger observations by the LV collaboration are expected. While the uncertainty for the tidal polarizability associated with GW170817 is not sufficient to constrain the EOS, this might change for future observations. For example, mergers with better signal-to-noise ratios could be observed, or sufficiently many mergers are observed so that accurate information can be extracted. In addition, third generation GW detectors might provide tidal-polarizability measurements with 10\% uncertainty.

To conclude, we pose the question if and when the accuracy of gravitational-wave observations will be sufficiently small to provide constraints on the EOS that are tighter than the ones from nuclear theory. From our results, we estimate that the uncertainty $\tilde{\Lambda}$ needs to be of the order of $\Delta\tilde{\Lambda}<300$ to test the chiral EFT prediction in the density range $n_\mathrm{sat}-2n_\mathrm{sat}$. Based on the contrast between MM and CSM, we expect that $\Delta\tilde{\Lambda}<100$ is needed to shed light on the possible existence of phase transitions in dense matter.

\section{Acknowledgement}
This work was supported in part by the U.S.~DOE under Grants 
No.~DE-AC52-06NA25396 and DE-FG02-00ER41132, by the LANL LDRD program, and by the National Science Foundation Grant No.~PHY-1430152 (JINA Center for the Evolution of the Elements). 
J.M. was partially supported by the IN2P3 Master Project MAC, "NewCompStar" COST Action MP1304, PHAROS COST Action MP16214.
This research used resources provided by the Los Alamos National
Laboratory Institutional Computing Program, which is supported by the
U.S. Department of Energy National Nuclear Security Administration under Contract No. 89233218CNA000001. Computational resources have been provided by the National Energy Research Scientific Computing Center (NERSC), which is supported by the U.S. Department of Energy, Office of Science, under Contract No. DE-AC02-05CH11231. Computational resources have also been provided by the J\"ulich
Supercomputing Center.



\end{document}